\documentclass[aps,prd,preprintnumbers,nofootinbib,twocolumn]{revtex4}
\usepackage{bm}
\usepackage{latexsym}
\usepackage{dcolumn}
\usepackage{amsmath,amsfonts,amssymb}
\usepackage{graphicx,epsfig}
\usepackage{psfrag}
\usepackage{amsthm}
\usepackage{color}
\usepackage{amsmath, amssymb, graphics, setspace}

\newcommand{\mathsym}[1]{{}}
\newcommand{\unicode}[1]{{}}

\usepackage{bm}
\usepackage{latexsym}
\usepackage{dcolumn}
\usepackage{amsmath,amsfonts,amssymb}
\usepackage{graphicx,epsfig}
\usepackage{psfrag}
\usepackage{amsthm}
\usepackage{color}

\usepackage{nccmath}
\usepackage{moresize}
\usepackage{enumerate}
\usepackage[hang,flushmargin]{footmisc}
\usepackage[titletoc,toc]{appendix}
\usepackage{lipsum}
\usepackage{subcaption}
\usepackage{epstopdf} 
\usepackage{hyperref}
\usepackage{rotating}
\usepackage{multirow}
\usepackage{mathtools}

\usepackage{stackengine}
\usepackage{scalerel}

\newcommand{\be}{\begin{equation}}
\newcommand{\ee}{\end{equation}}
\newcommand{\bea}{\begin{eqnarray}}
\newcommand{\eea}{\end{eqnarray}}
\newcommand{\bse}{\begin{subequations}}
\newcommand{\ese}{\end{subequations}}
\newcommand{\bce}{\begin{center}}
\newcommand{\ece}{\end{center}}
\newcommand{\bfg}{\begin{figure}}
\newcommand{\efg}{\end{figure}}
\newcommand{\bit}{\begin{itemize}}
\newcommand{\eit}{\end{itemize}}
\newcommand{\bed}{\begin{description}}
\newcommand{\eed}{\end{description}}
\newcommand{\ben}{\begin{enumerate}}
\newcommand{\een}{\end{enumerate}}
\newcommand{\nn}{\nonumber}

\newcommand{\fr}{\frac}
\newcommand{\sq}{\sqrt}
\newcommand{\no}{\noindent}

\def\le {\left}
\def\ri {\right}
\def\e  {\epsilon}

\def\l  {\lambda}
\def\L  {\Lambda}

\def\O  {\Omega}

\def\s  {\sigma}

\newcommand{\cR}{\mathcal R}

\newcommand{\OLp}{\O_{_0} \!\!^{\!(\!\L\!)}}

\newcommand{\ONp}{\O^{(NR)}_{_0}}
\newcommand{\Olp}{\O^{(L)}_{_0}}

\newcommand{\Hp}{H_{\text{\tiny 0}}}

\newcommand{\bdm}{\begin{displaymath}}
\newcommand{\edm}{\end{displaymath}}

\begin{document}

\title{Dark matter or strong gravity?}

\author{Saurya Das} \email[email: ]{saurya.das@uleth.ca}
%


\affiliation{Theoretical Physics Group,
Department of Physics and Astronomy,
University of Lethbridge, 4401 University Drive,
Lethbridge, Alberta T1K 3M4, Canada}

\author{Sourav Sur}
\email[email: ]{sourav.sur@gmail.com}

\affiliation{Department of Physics and Astrophysics, University of Delhi, New Delhi 110007, India}

\begin{abstract}
%

We show that Newton's gravitational potential,
augmented by a logarithmic term,
partly or wholly mitigates the need for dark matter. 
As a bonus, it also explains why MOND seems to work at galactic scales. We speculate on the origin of such a potential.

\vspace{0.3cm}
\noindent
%

\noindent
{\bf This essay received an Honorable Mention in the 2022 Gravity Research Foundation Essay Competition}\\

\end{abstract}

\maketitle


It is well known that the flatness of galaxy rotation curves, gravitational lensing from galaxies, and the $\Lambda$CDM model of cosmology point towards the existence of a significant amount (about 25\%) of invisible matter in the Universe, also known as Dark Matter (DM). 
Yet, despite there being no shortage of viable theories, the true nature of DM remains elusive as ever \cite{dm1}. This suggests that while one waits patiently for evidences of DM to emerge from a number of current and future experiments \cite{dm2,dm3}, 
it is important to explore alternative theories that can potentially 
explain observations without the need for DM. 
In this essay, we explore one such mechanism and 
show that a simple addition of
a logarithmic term to the Newton's gravitational potential can precisely do that, 
and thereby obviate the need for DM, or at least reduce its importance in astrophysics and cosmology. 
Before going into the details, we first argue that this is indeed a viable proposition. 
Note that, 
to replace DM, one needs `more gravity' than the attractive $1/r$ potential of Newton provides. 
Second, it must not significantly alter the Newtonian potential at `short' distances, that is, at sub-galactic scales. 
These two conditions require any additional potential to go slower than 
$1/r$ which leads to the natural choice of a logarithmic potential. 
In the language of fields, this means that 
in addition to the standard Newtonian $-1/r^2$ 
field, there is now also a $-1/r$ field. 
Combining the two terms, one can now write the `total' gravitational potential and force on a particle of mass $m$ due to another mass $M$ separated by a distance $r$,
respectively as 
\begin{eqnarray}
V(r) &&= - \frac{GMm}{r} \,
+ \lambda\,M m\,\ln \le(\frac{r}{r_0} \ri) 
\label{pot1} \\
F(r) &&= - \frac{GMm}{r^2} \,-\, 
\frac{\lambda Mm}{r}~, \label{force2}
\end{eqnarray}
where $\lambda$ and $r_0$ are constants, the latter having no observational consequence. 
For the logarithmic 
term to start dominating at galactic length scales $r_1 \approx 10^{21}$ m  $\approx 30$ kpc, 
i.e.,
setting the two terms in Eq.(\ref{force2}) to be approximately equal, it follows that $\lambda=G/r_1$. 
The new $1/r$ term in Eq.(\ref{force2}) immediately gives rise to flat galaxy rotation curves, as can be seen by equating it to the centrifugal force 
on a particle of mass $m$ in a rotation curve
by a galaxy of mass $M$
%
%
\begin{eqnarray}
&& \frac{\lambda Mm}{r} = \frac{m v^2}{r} \nonumber \\
\Rightarrow~&& v = 
\sqrt{\lambda\,M}
=\sqrt{\frac{GM}{r_1} }~,  \nonumber 
\end{eqnarray}
which is a constant for a given $M$. 
Plugging in $r_1=30$ kpc and $M=10^{42}$ kg, 
one gets $v\approx 10^5$ m/s, which is
the typical speed of the spiral arms. 

To show that the logarithmic 
potential explains other phenomena at the galactic scales and beyond, we now consider the gravitational lensing of light from a distant star by a galaxy (or a galaxy cluster). 
The total deflection angle, for an impact parameter $b$, is now given by \cite{carroll}
\begin{eqnarray}
\delta &=& \fr 2 {c^2} \int \vec\nabla_\perp 
V ds \nn \\ 
&=& 
\frac{2 G M b}{c^2} \int_{-\infty}^{\infty} \fr{dx}
{(b^2+x^2)^{3/2}}
+ \fr{2\l\,M b}{c^2} \int_{-\infty}^\infty 
\fr{dx}{(b^2+x^2)} \nn \\
&=& \fr{4GM}{b c^2} +\, \fr{2 \l \, M \pi}{c^2} 
\equiv \fr{4GM'}{b c^2} 
%
\end{eqnarray} 
where $M'=M(1+b\pi/2r_1)$.
Thus, the estimated galaxy mass $M'$ via lensing is greater than its actual mass $M$.
In other words, by including the logarithmic term, one sees that the observed lensing can be explained by a much smaller 
amount of galactic matter, thus reducing the need for DM, or eliminating it completely. Again, this is due to the extra gravity that this term provides. 
%
The proportional 
error in mass estimation by omitting the logarithmic term is given by 
$\epsilon\equiv (M'-M)/M={\lambda \, b \pi}/{2 G}=
{b \pi}/{2 r_1}$. 
Then, 
for a galaxy or a cluster of galaxies with 
$b \simeq 10$ to $100$ kpc 
(i.e., the average size of a galaxy or cluster), 
one obtains 
$
\epsilon \simeq\, 1~\text{to}~10 \,. 
$
That is, the estimated mass may be 
as much as $10$ times as large as
the actual mass, which is far from insignificant!
This applies to, for example, the estimation of the bullet cluster mass via lensing,
implying a potentially 
significant over-estimation of the amount of DM therein \cite{bullet1,bullet2}. 

\begin{figure}[!htp]
\includegraphics[height=3.25in,width=3.37in]{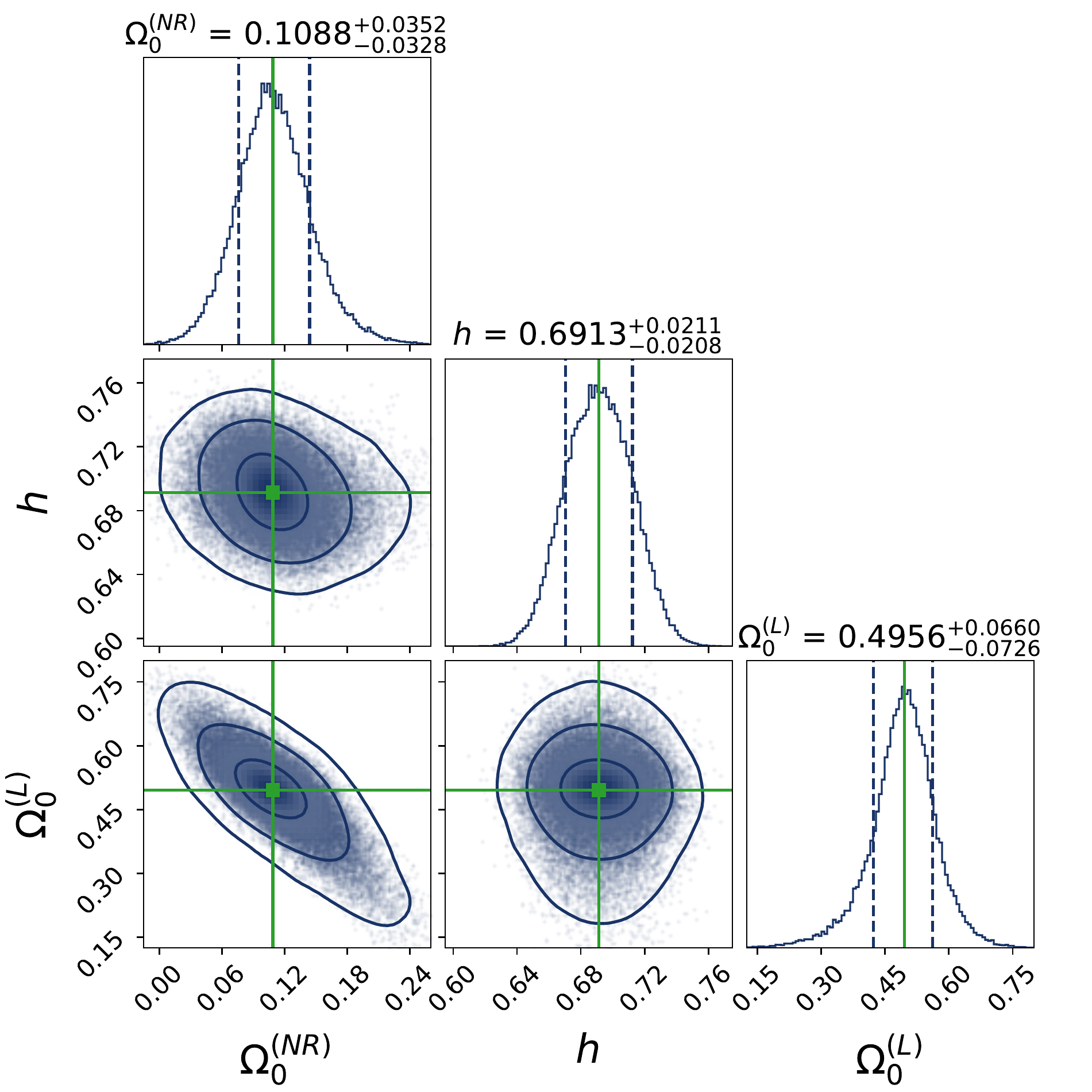}
\caption{\footnotesize The $1\s$-$3\s$ parametric contour levels obtained 
using the Pantheon (binned) data combined with the  
Hubble data. Here $h= \Hp/\!\le[100 \, \mbox{Km s}^{-1}\, \mbox{Mpc}^{-1}\ri]$ is the reduced Hubble constant.}
\label{LP_PanHubb_Fig}
\end{figure}

Next, let us turn to cosmology. 
It can be shown that the logarithmic term modifies the Friedmann equation 
as 
\cite{rai,dassur1}
\be
H^2 (z) = H_0^2 \le[\ONp \,f_{NR} (z) + \OLp \, f_{\Lambda} (z)
+\,\Olp \,f_{L} (z)  \ri]~,
\label{hub1}  
\ee 
%
where $z$ denotes the redshift, $H_0$ is the Hubble constant, i.e. the
value of the Hubble parameter $H(z)$ at the present epoch, $\ONp$, $\OLp$ and $\Olp$ are respectively the values of the density parameters 
corresponding to the
non-relativistic matter (dust), cosmological constant $\L$, and the
logarithmic term, and  
\begin{eqnarray}
f_{NR} (z) &&= (1+z)^{3} \nonumber  \\
f_{\Lambda} (z) &&= 1 ~ \nonumber \\
f_{L} (z) &&= (1+z)^{2} \ln (1+z) ~. 
\end{eqnarray}
Note that we have ignored the radiation term, and assumed a spatially flat universe. At $z=0$, Eq.(\ref{hub1}) gives 
$\ONp +\OLp = 1$. Hence eliminating $\OLp$ from Eq.(\ref{hub1}), we 
carry out
the statistical parametric estimation using the Supernovae type-Ia 
Pantheon (binned) data
\cite{scol,panth}
in combination with the observational Hubble $(H(z))$ data
\cite{MPCJM-OHD,YRW-OHD,RDR-OHD}.
The standard procedure for the Markov-chain Monte Carlo (MCMC) random probabilistic exploration
(see e.g. the section 4 of 
\cite{sdss_BEC_QB}),
leads to the parametric best fit and contour levels 
shown in Fig.\,\ref{LP_PanHubb_Fig}. We see that the best fit of $\ONp$ 
is significantly lower (only $\simeq 0.11$) compared to about $0.3$ for
$\L$CDM. So, if we take out the visible baryonic dust contribution, which
is $\simeq 0.05$, then we will be left with an effective CDM contribution
of about only $6\%$ of the total. This is significantly smaller than
the usual estimate of about $25\%$, and shows consistency with our starting hypothesis 
of a much reduced need for DM.

Having demonstrated that the simple addition of 
a logarithmic term to the standard Newtonian potential can explain multiple 
phenomena normally attributed to DM at the 
galactic and cosmological length scales, let us speculate on the origin of such a term. 
There are a couple of routes one can take. For example, one can simply posit it as a fundamental law of nature and try to embed it
into a covariant theory. 
The $f(R)$ theories of gravity theory 
may help, since it is known that 
the corresponding action can be transformed 
into a standard Einstein-Hilbert action coupled  minimally to a scalar field with a self-interacting potential
\cite{fr}. Therefore, working backwards,
in principle one can try to find 
an appropriate scalar potential that would resemble the matter term required for the logarithmic term, and in turn arrive at a suitable $f(R)$ model of gravity. 

The above approach is not free of problems, plus there exists a much more interesting alternative as follows. We propose that the potential in Eq.(\ref{pot1})
is nothing but the {\it quantum potential}, generated by an underlying wavefunction $\Psi = \cR\,e^{i\,S} \,,$ ($\cR,S \in \mathbb{R}$), as per 
\cite{bohm}
\be
V=V_Q =\, -\, \fr{\hbar^2}{2m} 
\frac{\nabla^2 \cR}{\cR} ~.
\label{qp1}
\ee
This possibility cannot be ruled out, since 
as shown in \cite{dassur0}, there is no mechanism for a particle to distinguish a background classical potential and its wavefunction dependent 
quantum potential. In fact, it sees the sum of the two. 
Eq.(\ref{qp1}) can be re-written as 
\be
\nabla^2 \cR  + \frac{2m\,V_Q}{\hbar^2}\,{\cal R} = 0~.
\ee
It can be shown that 
the above differential equation 
has an unique solution for the potential 
under consideration 
in terms of Bessel and Airy functions, with standard boundary conditions, namely 
\cite{dassur1}
%
%
\begin{eqnarray}
\cR &=& 
 \sqrt{\frac{\ell_1} r}\, I_1\left(2\,
\sqrt{\frac{r} {\ell_1}}\right),
\quad r\leq r_1 \,, 
\nonumber \\
&=&  \frac 1 r \, Ai \left( (-1)^{1/3}\,\left[ \fr{r_0}{\ell_2}\right]^{2/3}
{\left[\frac{r}{r_0} - 1\right]} 
\right) \nonumber \\
&+&  \frac 1 r \, Bi \left( (-1)^{1/3}\,\left[ \fr{r_0}{\ell_2}\right]^{2/3}
{\left[\frac{r}{r_0} - 1\right]} 
\right)
  + c.c. ~,~~ r \geq r_1 \nonumber 
\end{eqnarray}
where $\ell_1$ is a constant. 
Note that the phase of the wavefunction remains
undetermined. 

Before concluding, we note that 
Modified Newtonian Dynamics (MOND) can be re-written in terms of an effective 
logarithmic potential, but with 
$\lambda = \sqrt{G a_0/M}$ 
\cite{MOND1}, 
where $a_0 = 1.2\times 10^{-10}~m/s^2$. 
The above $\lambda$, although having a 
different functional form,
numerically equals our predicted value of 
$G/r_1$, for typical galaxy masses!
%
As a result, for gravitational lensing from galaxies 
with $M = \le(10^{12} - 10^{14}\ri) M_\odot$ 
(where $M_\odot$ is the solar mass), one has
$\e = 
{\pi b \sq{a_0}}/{\sq{GM}}=1$ to $10$.
This in turn explains why 
%
%
%
%
MOND correctly predicts the 
flat rotation curves without DM. 
We emphasize however, 
that unlike MOND, the current proposal is not designed to explain just one phenomenon at a given length scale, but can potentially explain a host of 
phenomena at {\it all} length scales. 

To conclude, in this essay, we have shown that the addition of a simple logarithmic potential to the standard Newtonian gravitational potential 
explains a number of astrophysical phenomena, which are normally attributed to DM. 
Thus at the very least, it reduces the importance of DM, and in the best case scenario, can 
mitigate its need altogether.
Furthermore, there can be more than one origin 
of this additional potential, either in the framework of extended gravity theories, or as an emergent phenomenon stemming from an underlying quantum wavefunction.
Either way, we believe that the proposal deserves to be explored further to tighten its theoretical underpinnings as well as to test its predictions with other astrophysical and cosmological observations.


\vspace{5pt}
\noindent 
{\bf Acknowledgment}

\no
This work was supported by the Natural Sciences and Engineering
Research Council of Canada.
%



\end{document}